\date{}
\documentclass[10pt,aps,prl,superscriptaddress,showpacs,twocolumn]{revtex4-1}
\usepackage{graphicx,psfrag,amsmath,amssymb,amsfonts,latexsym,color,dcolumn,bm}
\usepackage[hidelinks]{hyperref}

\begin{document}

\title{Topological quantum friction}

\author{M. Bel\' en Farias}
\affiliation{Departamento de F\'isica, FCEyN, UBA and IFIBA, CONICET,
Pabell\'on 1, Ciudad Universitaria, 1428 Buenos Aires, Argentina}
\affiliation{Center for Nonlinear Studies, MS B258, Los Alamos National Laboratory, Los Alamos, NM 87545, USA}
\affiliation {Theoretical Division, MS B213, Los Alamos National Laboratory, Los Alamos, NM 87545, USA}

\author{Wilton J. M. Kort-Kamp}
\affiliation{Center for Nonlinear Studies, MS B258, Los Alamos National Laboratory, Los Alamos, NM 87545, USA}
\affiliation {Theoretical Division, MS B213, Los Alamos National Laboratory, Los Alamos, NM 87545, USA}

\author{Diego A. R. Dalvit}
\affiliation {Theoretical Division, MS B213, Los Alamos National Laboratory, Los Alamos, NM 87545, USA}

\date{today}

\begin{abstract}  
We develop the theory of quantum friction in two-dimensional topological materials. 
The quantum drag force on a metallic nanoparticle moving above such systems is sensitive to the non-trivial topology of their electronic phases, shows a novel distance scaling law, and 
can be manipulated through doping or via the application of external fields. We use the developed framework to investigate quantum friction due to the quantum Hall effect in magnetic field biased graphene, and to topological phase transitions in the graphene family materials. It is shown that topologically non-trivial states in two-dimensional materials enable an increase of two orders of magnitude in the quantum drag force with respect to conventional neutral graphene systems.
\end{abstract}
\maketitle

Quantum vacuum fluctuations of the electromagnetic field produce observable macroscopic effects, the most renowned example being the attractive Casimir force between two neutral bodies \cite{Casimir1948,Milonni,Springer,RMP}.  
When the bodies are set into relative motion at constant velocity, a dissipative force that opposes the motion is exerted on each of them due to the exchange of Doppler-shifted virtual photons, an effect known as quantum friction \cite{Pendry1997,Volokitin2007}. 
Various theoretical studies have been carried out to model surface-surface (Casimir) (see, e.g., \cite{Barton2011,Hoye2014}) 
and particle-surface (Casimir-Polder) quantum friction  (see, e.g., 
\cite{Scheel2009,Intravaia2014,Intravaia2016}), analyzing the velocity and distance dependency of the drag force in 3D bulk materials and, more recently, in graphene \cite{Volokitin2016,Farias2017}. Due to its short range and small magnitude, measurements of quantum friction in mechanical moving systems are challenging, but the analog phenomenon of Coulomb drag \cite{Pogrebinskii1977, Price1983,Volokitin2011}, in which a current in one plate induces a voltage bias in another one via the fluctuating Coulomb field, has been successfully demonstrated in quantum wells as well as in graphene \cite{Gramila1991,Gorbachev2012}.

When the optoelectronic response of the bodies has non-trivial topological features, novel Casimir physics phenomena can arise 
due to the interplay between quantum vacuum fluctuations and topologically protected surface states. 
Quantized Casimir forces and spontaneous emission have been found in magnetic field biased graphene \cite{Tse2012,Kort-Kamp2015} and in Chern insulators \cite{Rodriguez-Lopez2014} due to the quantum Hall effect (QHE). 
More recently, Casimir force topological phase transitions (TPT)  \cite{Rodriguez-Lopez2017} have been predicted in the graphene family materials, formed by silicene, germanene, stanene, and plumbene \cite{Castellanos2016,Mannix2017,Molle2017}. The interplay between Dirac physics, spin-orbit coupling, and externally applied electrostatic and polarized laser fields can drive these materials through various topological phases \cite{Ezawa2013}, resulting in novel Casimir force distance scaling laws, abrupt magnitude changes, and also force quantization and repulsion. 

In this Letter, we study the impact of two-dimensional (2D) topological materials on quantum friction. We show that while the electric component of the Casimir-Polder frictional force is sensitive only to the non-topological longitudinal conductivity of the material, the magnetic component depends on the Hall conductivity and can hence probe topological features manifested through the charge Chern number of the monolayer. We exemplify these general findings by studying topological quantum friction on a metallic nanoparticle due to the QHE in magnetic field biased graphene and in TPT with the graphene family materials.


{\it Quantum friction in the flatland:}
Consider a nanoparticle moving with constant velocity ${\bf v}$ at a distance $d$ from a 2D topological material (see Fig.1). 
The optical response of the nanoparticle is assumed to be isotropic and given by its electric $\alpha_E(\omega)$ and magnetic $\alpha_H(\omega)$ polarizabilities, while the monolayer is characterized by a rotationally invariant conductivity tensor $\sigma_{ij}(\omega) = \sigma_L(\omega) \delta_{ij} +  \sigma_H(\omega) \varepsilon_{ij}$, ($i,j=x,y$), where $\sigma_L$ and $\sigma_H$ are the longitudinal and Hall conductivities, and $\varepsilon_{ij}$ is the 2D Levi-Civita tensor. We assume the motion is along the $\hat{\bf x}$ direction,
and that the particle's trajectory is prescribed by means of an external force ${\bf F}_{\rm ext}$ along that same direction.
To lowest order in velocity, the (zero temperature) quantum frictional force $\textbf{F}= - F \hat{\bf x}$ is the sum of an electric and magnetic contribution, $F= F_{E} + F_{H}$ \cite{Intravaia2014,Volokitin2008}
\begin{equation}
\label{eq:forcegral}
F_{E,H} = \frac{\hbar v_x^3}{12 \pi^3} {\alpha}^{\prime}_{E,H;I}(0) \!\! 
\int_{-\infty}^{\infty} \! \!  \! \! \!\!\! dq_y \!\! \int_0^{\infty} \! \! \! \! \!\!\! dq_x q_x^4 
{\rm Tr} \! \left[\underline{G}_{E,H;I}^{\prime}({\bf q},d,0) \right] ,
\end{equation}
where the subscript $I$ denotes imaginary part, the primed superscript means derivative with respect to frequency, and $\underline{G}_{E,H}$ is the scattered part of the electric/magnetic Green tensor of the 2D sheet \cite{Green}. 
For distances $d\gg v_F \tau$, where $v_F$ is the Fermi velocity and $\tau$ is the electronic relaxation time of the involved materials, corrections to Eq.(\ref{eq:forcegral}) due to spatial dispersion effects can be neglected \cite{Reiche2017}.
Note that quantum friction is a low-frequency phenomenon and can therefore probe the static topological response of the monolayer, as we will show below.

\begin{figure}
\centering
\includegraphics[width=\linewidth]{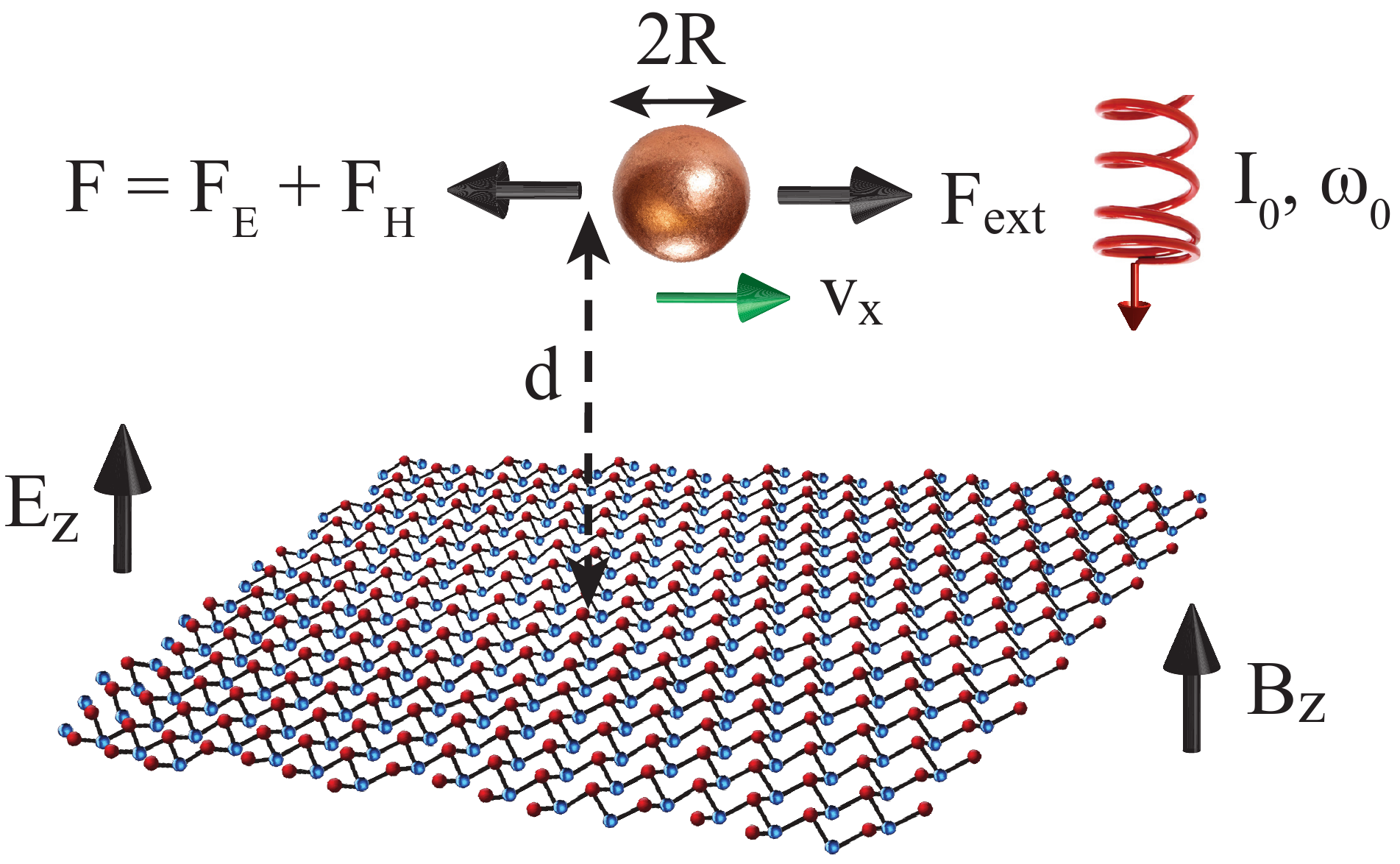}
\caption{Topological quantum friction in the flatland. A metallic nanoparticle moves parallel to a 2D topological material. Examples considered in this work are monolayers of the graphene family in the presence of a static magnetic field, a static electric field, or a circularly polarized laser. 
} 
\label{Fig1}
\end{figure}

Analytical expressions for the drag forces can be found in the near-field regime, where they are strongly enhanced.
In this case one finds ${\rm Tr} \left[\underline{G}'_E({\bf q},d,\omega) \right] =  (q/2) e^{-2 q d} [2 r'_{pp}+ (2 \omega/c^2 q^2) r_{ss} + (\omega^2/c^2 q^2)  r'_{ss}]$, where
the diagonal Fresnel reflection coefficients for the 2D material are given by 
$r_{ss} \approx  - \mu_0 \omega (\sigma_{L}^2+\sigma_{H}^2)/\mathcal{D}$ and 
$r_{pp} \approx [ 2 i q \sigma_{L} +  \mu_0 \omega (\sigma_{L}^2+\sigma_{H}^2) ] / \mathcal{D}$, with 
$\mathcal{D}= 4  \epsilon_0 \omega  + 2i q \sigma_{L}  +  \mu_0 \omega  (\sigma_{L}^2+\sigma_{H}^2)$ \cite{Kort-Kamp2015} and $q=|{\bf q}|$.
 A similar expression for 
${\rm Tr} \left[ \underline{G}'_H \right]$ can be found using electromagnetic duality, which amounts to 
swapping $s$ and $p$ in the previous expressions. For a spherical metallic nanoparticle, 
$\alpha_E(\omega)= 4 \pi R^3 [\epsilon(\omega)-1]/[\epsilon(\omega)+2]$ and $\alpha_H(\omega)= (2 \pi/15)  R^3 (\omega R/c)^2 [\epsilon(\omega)-1]$,
where $R$ is the radius of the particle and $\epsilon(\omega)=1-\omega_p^2/(\omega^2 + i \gamma \omega)$ is the Drude permittivity of  the constituent material  \cite{Bohren}. Using the above expressions in Eq.(\ref{eq:forcegral}), we obtain
\begin{eqnarray}
F_{E} &=&  \frac{45 \epsilon_0  R^3 \hbar \gamma}
{32 \pi \omega_p^2 }  \; \frac{1}{\sigma_{L}(0)} \; \frac{v_x^3}{d^6}, 
\label{eq:forcesimpleE}
\\
F_{H} &=& \frac{\mu_0 R^5 \hbar \omega_p^2}
{256 \pi c^2 \gamma} \; \frac{\sigma_{L}^2(0)+\sigma_{H}^2(0)}{\sigma_{L}(0)}  \; \frac{v_x^3}{d^6}.
\label{eq:forcesimpleH}
\end{eqnarray}
If instead of a metallic nanoparticle one considers a dielectric one, then $\alpha'_{H,I}(0)=0$ and the frictional force is solely given by the electric component. Also, for 2D systems with non-rotational invariant conductivity tensor (e.g., phosphorene), the drag forces are still given by the equations above, replacing $\sigma_L^2(0) \rightarrow \sigma_L(0) \sigma_T(0)$ in Eq.(\ref{eq:forcesimpleH}), where $\sigma_T$ is the transverse component of the conductivity tensor. 
Due to the 2D nature of the plate, 
both $F_E$ and $F_H$ have the same $d^{-6}$ distance scaling law and 
can be of the same order of magnitude. This is in stark contrast with quantum friction
between a nanoparticle and a 3D bulk, for which $F_{E;B} \propto d^{-7}$,
$F_{H;B} \propto d^{-5}$ and $F_{E;B} \ll F_{H;B}$ for good conductors \cite{Bulk} (see Fig. 2). 


\begin{figure}
\centering
\includegraphics[width=0.9\linewidth]{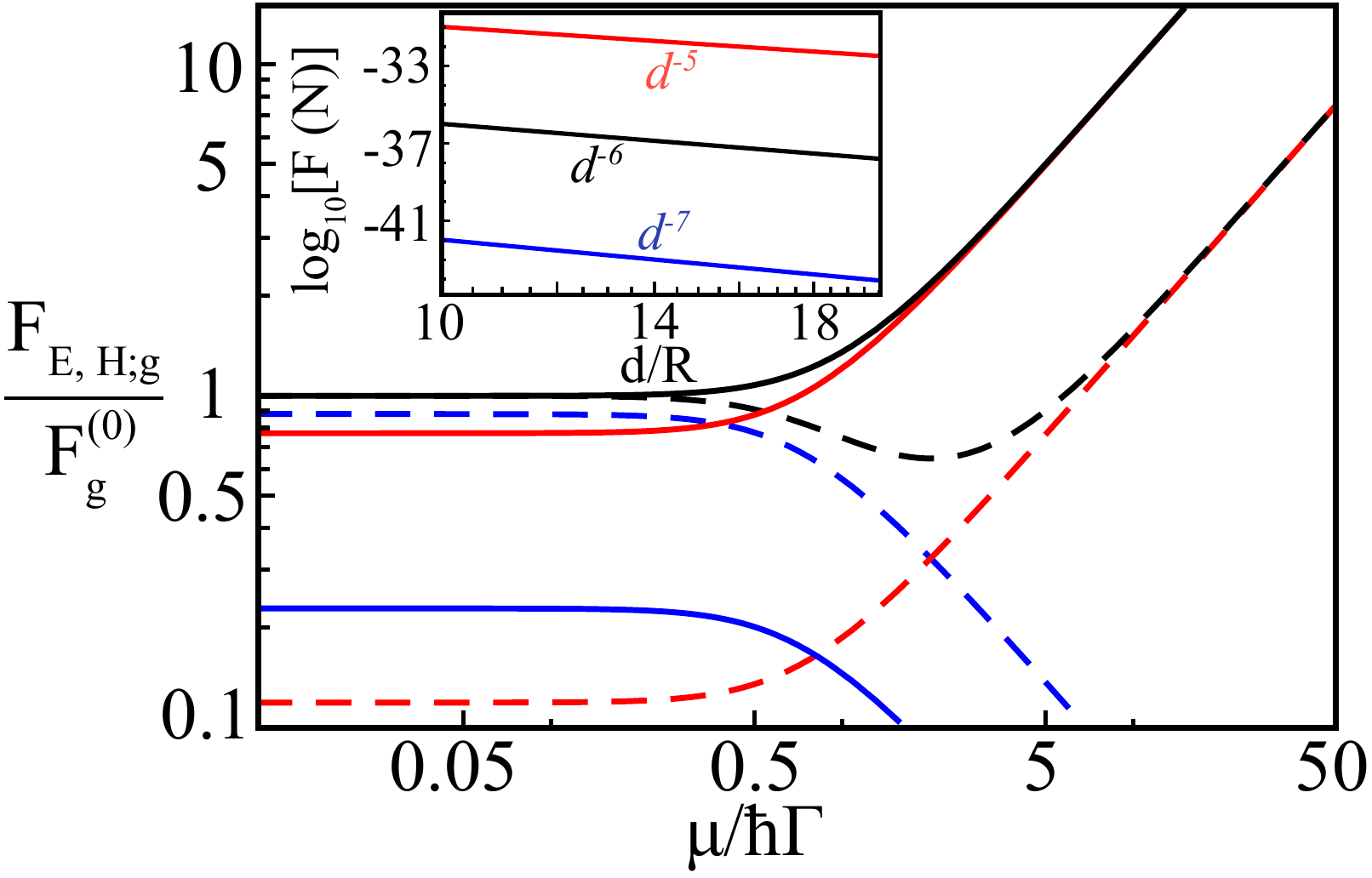}
\caption{
Electric (blue), magnetic (red), and total (black) quantum frictional forces  
due to unbiased graphene as a function of doping for different values of $R=50$ nm (solid) and 10 nm (dashed).
The normalization is the total drag force $F^{(0)}_g= F^{(0)}_{E;g}+F^{(0)}_{H;g}$ for unbiased undoped graphene. 
Inset: Quantum friction versus distance for a nanoparticle moving above graphene (black) or a 3D metallic bulk.
For the latter we separately show the electric (blue) $F_{E;B}$ and magnetic (red) $F_{H;B}$ 
components. Parameters are $R=50$ nm, $v_x=340$ m/s, and copper ($\hbar \omega_p= 7.4$ eV and $\hbar \gamma=9.1$ meV \cite{Ordal}) is the constituent material of both the nanoparticle and the bulk.
}
\label{Fig2}
\end{figure}


The most important feature of the above equations is the dependency of the magnetic force on the Hall conductivity. According to the Thouless, Kohmoto, Nightingale, Nijs (TKNN) theorem \cite{Thouless1982}, for insulating phases $\sigma_H(0)= (e^2/2 \pi \hbar) C$,
where $C=-i \sum_{\beta} \int d^2{\bf k} (2 \pi)^{-1} \hat{\bf z} \cdot \nabla_{{\bf k}} \times \langle u_{{\bf k}}^{\beta} |  \nabla_{{\bf k}} u_{{\bf k}}^{\beta} \rangle$ is the charge Chern number (a topological invariant), the sum is over occupied electron sub-bands $\beta$, the ${\bf k}$-integral is over the first Brillouin zone, and $u_{{\bf k}}^{\beta}$ are the eigenfunctions of the Hamiltonian of the monolayer. The corresponding longitudinal conductivity 
can be derived from Kubo's formula
\cite{Kubo1957}, resulting in
$\sigma_L(0) = e^2 \Gamma/ 2 \pi E_{\Gamma}$, where
$E^{-1}_{\Gamma}= \sum_{\beta\beta'} \int d^2{\bf k} (2 \pi)^{-2} 
(\varepsilon^{\beta'}_{{\bf k}} \!\!\!-\! \varepsilon^{\beta}_{{\bf k}}) \; |\langle u_{{\bf k}}^{\beta'} |  \nabla_{{\bf k}} u_{{\bf k}}^{\beta} \rangle   |^2 / [(\varepsilon^{\beta'}_{{\bf k}} \!\!\!-\! \varepsilon^{\beta}_{{\bf k}})^2 + \hbar^2 \Gamma^2]$, the $\beta'$ sum is over unoccupied sub-bands, $\varepsilon^{\beta(\beta')}_{{\bf k}}$
are the eigen-energies corresponding to the eigen-vectors $u_{{\bf k}}^{\beta(\beta')}$,
and $\Gamma$ is the electron scattering rate. For insulating phases with trivial topology ($C=0$), $F_E$ and $F_H$ have opposite behavior with $\sigma_L(0)$, the former (latter) increasing (decreasing) as the resistivity of the material grows. For non-trivial topology 
($C \neq 0$), and for small dissipation $\hbar \Gamma/ E_{\Gamma} \ll C^2$, both forces have the same behavior with $\sigma_L(0)$ and 
$F_H \propto C^2$ allows to probe the topology 
of the 2D material.



\begin{figure}[t]
\centering
\includegraphics[width=0.9\linewidth]{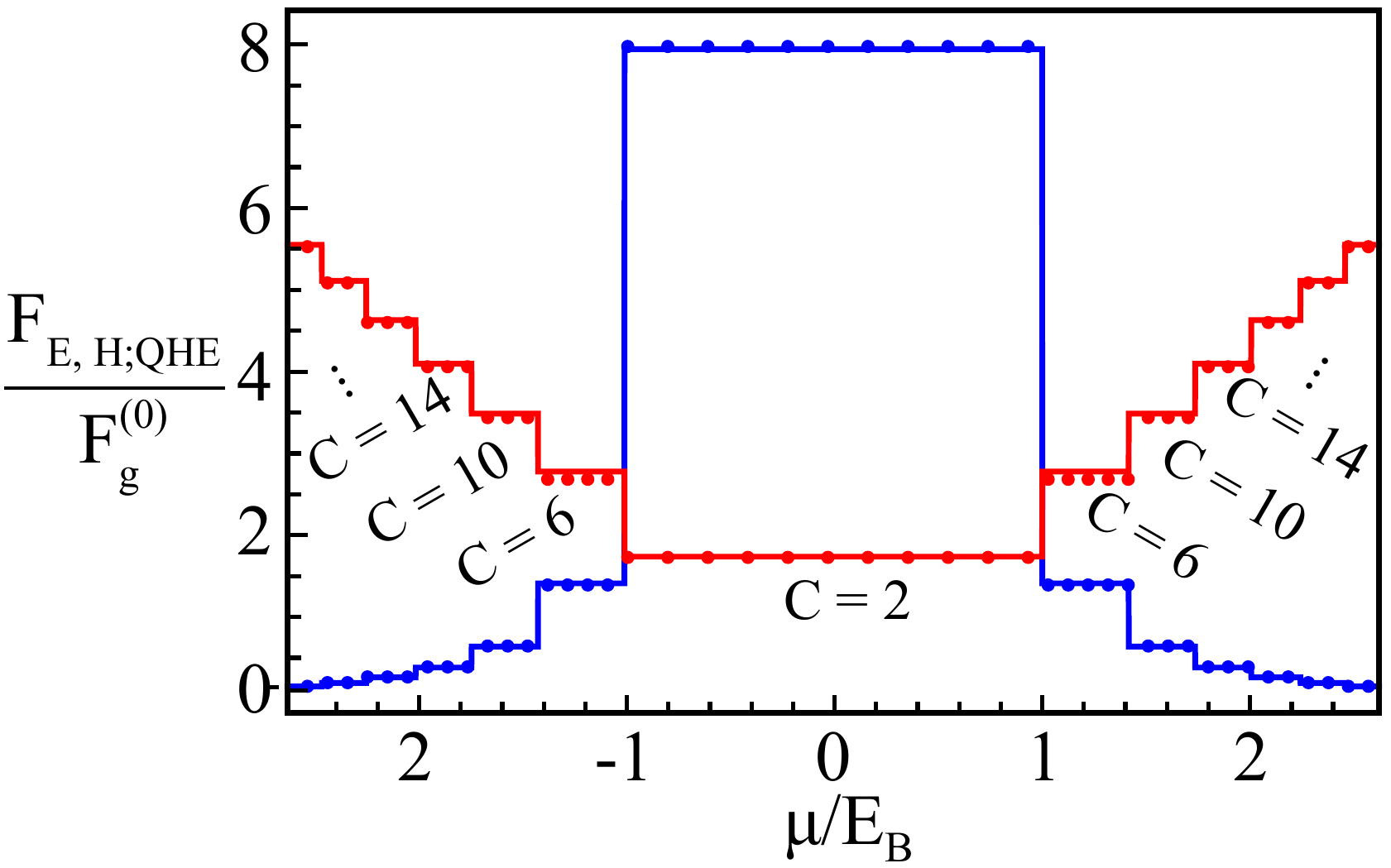}
\caption{Topological quantum friction on a nanoparticle due to the quantum Hall effect in graphene. Electric (blue) and magnetic (red) frictional forces as a function of doping. Solid lines correspond to Eqs.(\ref{qfrictionQHE}), and dotted lines to  
Eqs.(\ref{eq:forcesimpleE},\ref{eq:forcesimpleH}) where the exact formulas for $\sigma_{L,H}(0)$ are used \cite{Gusynin2007,Ferreira2012}. The Chern numbers of the magnetic plateaus are shown. Parameters are $B_z=10$ T, $\hbar \Gamma=8$ meV, and the Cu nanoparticle has radius $R=10$ nm. Due to the weak magnetic response of Cu, we can neglect effects of the magnetic field on the nanoparticle. 
} 
\label{Fig3}
\end{figure}


{\it Quantum friction and QHE in graphene:}
In order to introduce a normalization scale for the frictional forces, we first consider the simplest case of neutral unbiased graphene, that behaves as a semi-metal and has trivial topology. The corresponding expressions for the electric $F^{(0)}_{E;g}$ and magnetic $F^{(0)}_{H;g}$ frictional forces easily follow from Eqs.(\ref{eq:forcesimpleE},\ref{eq:forcesimpleH}) using that, in this case, $\sigma_H(0)=0$ and $\sigma_L(0)=\sigma_0$, where $\sigma_0=e^2/4 \hbar$ is graphene's universal conductivity. 
In Fig. 2 we show the effect of the chemical potential $\mu$ on $F_{E;g}$ and $F_{H;g}$, decreasing the former and increasing the latter.  For $\mu/\hbar \Gamma \ll 1$, $F_{E;g}$ dominates over $F_{H;g}$
for $R \ll \epsilon_0 c^2 \sqrt{360} \Gamma/\omega_p^2 \sigma_0$ (see dashed curves).  For large doping, 
$F_{E;g}/F^{(0)}_{g} \ll F_{H;g}/F^{(0)}_{g} \approx (4 \mu/\pi \hbar \Gamma) \times (1+F_{E;g}^{(0)}/F_{H;g}^{(0)})^{-1} $, 
which can lead to $\sim 25$ times enhancement of quantum friction over the total undoped force 
$F^{(0)}_{g} =F^{(0)}_{E;g}+F^{(0)}_{H;g}$ for
a $R=50$ nm Cu nanoparticle and typical graphene parameters $\mu=0.2$ eV and $\hbar \Gamma=8$ meV
\cite{Peresbook}.

We now study the impact of the QHE on quantum friction by considering that the graphene monolayer is subjected to a static perpendicular magnetic field $B_z$. 
When $B_z$ is strong enough that quantum Hall plateaus are well formed, and in the low dissipation limit
$\hbar \Gamma /E_B \ll \sqrt{N_c+1} - \sqrt{N_c}$
(where $N_c={\rm Floor}[\mu^2/E_B^2]$ and $E_B=\sqrt{2 e \hbar v_F^2 B_z}$), the DC Hall conductivity is given by $\sigma_H(0) \approx - (e^2/2 \pi \hbar)  (4N_c +2)$, that results from the addition of all allowed intra- and inter-band transitions
\cite{Gusynin2007,Ferreira2012}.
On the other hand, the static longitudinal conductivity is dominated by intra-band transitions for
$N_c \geq 1$, and takes the simple form  $\sigma_L(0) \approx (e^2 \Gamma/2 \pi) (1+\delta_{N_c,0})  (\sqrt{N_c+1}+\sqrt{N_c})^3 / E_B$. For $N_c=0$ inter-band transitions result in a correction to the longitudinal conductivity given by  $\approx e^2 \Gamma/4 \pi E_B$. We get
\begin{equation}
\frac{F_{E,H;{\rm QHE}}}{F^{(0)}_{E,H;g}} \! \approx \! \frac{E_B (2 + 3 \delta_{N_c,0})}{\pi \hbar \Gamma (\sqrt{N_c+1}+\sqrt{N_c})^3}
\!\! \times \!\!
\begin{cases}
\! \pi^2/4 & \!\!\!\!\! {\rm for \; E}, \\
\! (4N_c+2)^2 & \!\!\!\!\!{\rm for \; H}. 
\end{cases}
\label{qfrictionQHE}
\end{equation}
For fixed $B_z$ and varying $\mu$, both forces are quantized, depicting flat Hall plateaus between consecutive values of $N_c$, with jumps in the magnetic component depending on the QHE  topological invariant $C=4N_c +2$. In Fig. 3 we plot the quantum frictional forces as a function of doping, showing an excellent agreement between the approximated expressions given by Eq.(\ref{qfrictionQHE}) and the exact ones Eqs.(\ref{eq:forcesimpleE},\ref{eq:forcesimpleH}), in which we use the full (and cumbersome) expressions for the static conductivity tensor derived from Kubo's formula in \cite{Gusynin2007,Ferreira2012}. 
Although in the QHE regime $\sigma_L(0)$ contributes in the same way to both frictional forces, making them decrease as $\mu$ grows, the $(4 N_c+2)^2$ factor in $F_{H;{\rm QHE}}$ compensates that decrease and results in an overall growth of the magnetic quantum frictional force. If instead of fixing $B_z$ we fix $\mu$, then both forces increase as  $\sqrt{B_z}$ within any given plateau.

{\it Quantum friction and TPT in the graphene family:} 
2D hexagonal allotropes of Si, Ge, and Sn, namely silicene, germanene, and stanene, 
have been recently synthesized, enlarging the graphene family and bringing about a richer electronic structure \cite{Castellanos2016,Mannix2017,Molle2017}. 
They are staggered with finite buckling $2 \ell$, 
their spin-orbit coupling $\lambda_{SO}$ is non-zero, and their four Dirac cones can be controlled through the application of an external electrostatic field $E_z$ perpendicular to the layer, as well as
via an applied circularly polarized laser whose coupling to the layer is characterized by $\Lambda=\pm 8 \pi \alpha v_F^2 I_0/\omega_0^3$ ($\pm$ denotes left and right polarization, $\alpha$ is the fine structure constant, $I_0$ is the intensity of the laser, and $\omega_0$ is its frequency). 
The low-energy Dirac-like Hamiltonian per cone is given by $H_s^{\eta}=\hbar v_F (\eta k_x \tau_x + k_y \tau_y) +  \Delta_s^{\eta} \tau_z$,  where 
$\Delta_s^{\eta} = \eta s \lambda_{SO} - e \ell E_z - \eta \Lambda$ is half the mass gap. Here, $\eta,s= \pm 1$ are the valley and spin indexes, $k_{x,y}$ are in-plane components of the 2D wave vector, and $\tau_i$ are the Pauli matrices. This Hamiltonian is valid as long as $\omega_0$ is much greater than the hopping energy $t$ in the materials (typically in the range  $t \approx 1-3$ eV \cite{Ezawa2013}), and then the nanoparticle is almost transparent at such high frequencies. Also, the interaction between $E_z$ and the nanoparticle generates a spatially-dependent induced electric field $\sim E_z (R/d)^3$ on the monolayer, which can be neglected provided $R/d \ll1$. In summary, under these conditions the presence of the nanoparticle does not affect the coupling between the  monolayer and the external fields. These drive the 2D material through various electronic phases characterized by a charge Chern number $C = \frac{1}{2}\sum^{\star}_{s,\eta}\eta \; \text{sign}[\Delta_{s}^{\eta}]$ (see Figs. 4(a-b)),  where the star in the summations indicates that only terms with open gaps $\Delta^{\eta}_s \neq 0$ should be included. 

In the small dissipation limit $\hbar \Gamma \ll (\sum^{\star}_{\eta,s} |\Delta^{\eta}_s|^{-1})^{-1} \equiv 
\tilde{\Delta}$, and for neutral materials (where only inter-band transitions contribute to $\sigma_{L,H}$), one gets $\sigma_{H}(0)/\sigma_0 \approx (2/\pi) C$ and
$\sigma_{L}(0)/\sigma_0 \approx \hbar \Gamma/3 \pi \tilde{\Delta} +n_c/4$,  where $n_c$ is the number of closed gaps which accounts for the overlap between valence and conduction bands in semi-metallic phases  \cite{Tse2010,Nicol2012,KortKamp2017}.  
The resulting frictional forces are  
\begin{equation}
\frac{F_{E,H;{\rm TPT}}}{F^{(0)}_{E,H;g}} \approx \frac{1}{\frac{\hbar \Gamma}{3 \pi \tilde{\Delta}} + \frac{n_c}{4}}
\times
\begin{cases}
1 & {\rm for \; E}, \\
\left(\frac{\hbar \Gamma}{3 \pi \tilde{\Delta}} + \frac{n_c}{4}\right)^2 + \left(\frac{2C}{\pi}\right)^2 & {\rm for \; H}.
\end{cases}
\label{TPT}
\end{equation}
We first consider gapping neutral graphene with the external polarized laser, for which $n_c=0$, $C=-2 \; \text{sign}[\Lambda]$, and $\tilde{\Delta}= |\Lambda|/4$. For $\hbar \Gamma \ll |\Lambda|$, we obtain 
$F^g_{E;{\rm TPT}}/F_{E;g}^{(0)} \approx 3 \pi |\Lambda|/4 \hbar \Gamma$ and
$F^g_{H;{\rm TPT}}/F_{H;g}^{(0)} \approx (3 |\Lambda|/ \pi \hbar \Gamma) C^2$, so
both forces are enhanced with respect to the ungapped case.


\begin{figure}[t]
\centering
\includegraphics[width=\linewidth]{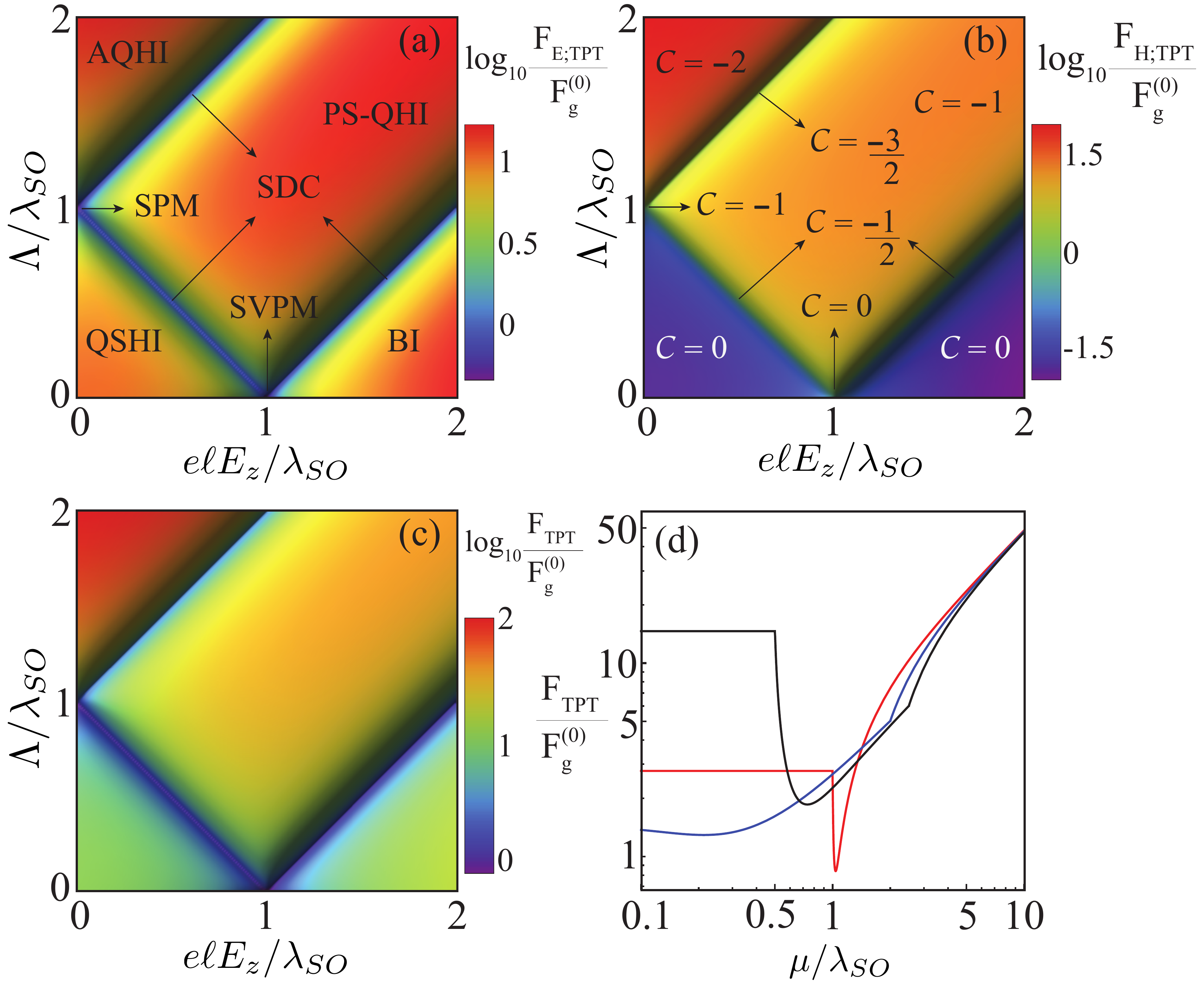}
\caption{Topological phase transitions in quantum friction. The electric (a), magnetic (b), and total (c) quantum frictional forces phase diagram for neutral graphene family materials. Acronyms for the phases are defined in the text. (d) Total force versus doping for selected points in phase space: $(e \ell E_z/\lambda_{\rm SO}, \Lambda/\lambda_{\rm SO})=$
(0,0) in red, (0,1) in blue, and (0,1.5) in black. All forces are normalized by $F_g^{(0)}$, $\hbar \Gamma / \lambda_{SO}=0.1$, Cu is used for the nanoparticle, and $R=50$ nm.} 
\label{Fig4}
\end{figure}

More interesting situations involving TPT occur for the other members of the graphene family. Figs. 4(a,b) show contour plots of 
$F_{E;{\rm TPT}}$ and $F_{H;{\rm TPT}}$ in the $(E_z, \Lambda)$ plane. For insulating phases with trivial topology, namely the quantum spin Hall insulator (QSHI) and band insulator (BI), the electric (magnetic) force is strongly enhanced (suppressed) since all mass gaps are non-zero, $F^{{\rm QSHI,BI}}_{E,H;{\rm TPT}}/F_{E,H;g}^{(0)} \approx (3 \pi \tilde{\Delta}/\hbar \Gamma)^{\pm 1}$. The frictional forces for the semi-metallic spin-valley polarized metal (SVPM) phase are simply one half (for the electric) or twice (for the magnetic) the force for unbiased undoped graphene, since the number of closed gaps is $n_c=2$ rather than 4. For insulating phases with non-trivial topology, namely the anomalous quantum Hall insulator (AQHI) and the polarized-spin quantum Hall insulator (PS-QHI), the electric force is the same as that for QSHI and BI phases (since it does not probe topology), while the magnetic one is given by
$F^{{\rm AQHI,PS-QHI}}_{H;{\rm TPT}}/F_{H;g}^{(0)} \approx  (12 \tilde{\Delta} / \pi \hbar \Gamma) C^2$. Unlike the insulating phases with trivial topology, the electric and magnetic forces for AQHI and PS-QHI phases increase as the effective gap $\tilde{\Delta}$ grows. The remaining spin-polarized metal (SPM) and single Dirac cone (SDC) phases have non-trivial topology and closed gaps ($n_c=2$ and 1, respectively). Similarly to the SVPM phase, the electric force is that of graphene weighted by $n_c$. In contrast to all other cases previously considered in this work, $\sigma_L(0)$ in the numerator of Eq.(\ref{eq:forcesimpleH}) cannot be neglected even in the limit of small dissipation, because of the contribution stemming from closed gaps (see Eq.(\ref{TPT})). Hence, the magnetic component of the frictional force shows an interplay between the topology resulting from the insulating behavior of cones with open gaps and the semi-metallic behavior from cones with closed gaps, namely 
$F^{{\rm SPM,SDC}}_{H;{\rm TPT}}/F_{H;g}^{(0)} \approx n_c/4 + 16 C^2/\pi^2 n_c$. Fig. 4(c) depicts the total drag force $F_{\rm TPT}$, which is dominated by its electric component in the QSHI/BI phases and by its magnetic one in the AQHI phase (where $F^{\rm AQHI}_{\rm TPT} \gtrsim 100  F^{(0)}_g$). Along the SDC phases the interplay between $n_c$ and $C$ can be clearly observed, with the one with $C=-3/2$  being approximately 3 times larger than the other two whose $C=-1/2$. 
Finally, we show in Fig. 4(d) the effect of doping on the total frictional force. For insulating phases where  $|\mu| < |\Delta^{\eta}_s|$ for all cones (black and red curves), $F_{\rm TPT}$ is unaffected by $\mu$. Once $|\mu|$ becomes larger than $|\Delta^{\eta}_s|$ for any given cone, intra-band transitions must be taken into account \cite{Tse2010,KortKamp2017}, and $F_{\rm TPT}$ initially decreases since $F_{E;{\rm TPT}}$ is  suppressed, and then increases because the growth of $F_{H;{\rm TPT}}$ overwhelms $F_{E;{\rm TPT}}$.  
In contrast, for semi-metallic phases $|\mu| > |\Delta^{\eta}_s|$ for at least one cone (blue curve), and $F_{\rm TPT}$ shows a smooth behavior as a function of doping. For large doping $|\mu| \gg \lambda_{\rm SO}$, all phases have the same limiting behavior $F_{\rm TPT} \approx F_{H;g}^{(0)} \times 4 |\mu| / \pi \hbar \Gamma$.

In summary, we developed the framework of topological quantum friction in 2D materials. 
Taking advantage of the TKNN theorem and the fact that quantum friction is a low-frequency phenomenon, we discovered that the Casimir-Polder frictional drag is sensitive to the underlying topology of monolayers supporting quantum Hall states. We also found that quantum friction satisfies a universal $d^{-6}$ distance scaling law, both for its electric and magnetic components and irrespective of the opto-electronic response of the monolayer. Casimir quantum friction between topological 2D materials can unveil an even richer phenomenology, e.g. an interplay between different Chern numbers corresponding to distinct phases. Topological quantum friction could in principle be measured in analog Coulomb drag experiments,
or in even more challenging mechanical quantum friction set-ups based on cryogenic atomic force microscopy. 
 

\begin{acknowledgments} 
We thank F. Intravaia, P. Ledwith, P. Rodriguez-L\'opez, and L. Woods for discussions. Financial support has been provided by the Los Alamos National Laboratory (LANL) Laboratory Directed Research and Development (LDRD) program, Center for Nonlinear Studies (CNLS), Agencia Nacional de Promoci\' on Cient\'ifica y Tecnol\'ogica (ANPCyT), Consejo Nacional de Investigaciones Cient\'ificas y T\'ecnicas  (CONICET), and Universidad de Buenos Aires (UBA).
\end{acknowledgments}



\begin{thebibliography}{1}
\bibitem{Bohren}
The expressions for the polarizabilities of the nanoparticle are valid as long as $R \ll c/\omega, c/\omega n(\omega)$, where $n(\omega)$ is the refractive index. See C. F. Bohren and D. R. Huffman,
{\it Absorption and scattering of light by small particles}
(Wyley-VCH, Weinheim, 2004).
\bibitem{Bulk}
The electric and magnetic quantum frictional forces between a metallic nanoparticle and a 3D metallic bulk are given by $F_{E;B} = 135 R^3 \hbar \gamma \gamma_B v_x^3/ 32 \pi \omega_p^2  \omega_{p;B}^2 d^{7}$ and $F_{H;B} = R^5 \hbar \omega_p^2 \omega_{p;B}^2 v_x^3/1280 \pi c^4 \gamma \gamma_B d^{5}$,
$\omega_{p;B}$ and $\gamma_B$ are the bulk plasma frequency and dissipation rate.
\end{thebibliography}
\end{document}